\documentclass{aa}
\usepackage{amssymb}
\begin{document}
\thesaurus{08(08.02.3; 08.16.5; 08.19.1)}
\title{Binary fraction in low-mass star forming regions:
  a reexamination of the possible excesses and implications}
\titlerunning{Binary fraction in low-mass star forming regions}
\author{G. Duch\^ene} 
\mail{Gaspard.Duchene@obs.ujf-grenoble.fr} 
\institute{Laboratoire d'Astrophysique, Observatoire de Grenoble,
  Universit\'e Joseph Fourier, BP 53, 38041 Grenoble Cedex 9, France}
\date{Received 23 July 1998; Accepted 28 October 1998}
\maketitle

\begin{abstract}
  
  Various surveys of low-mass binaries in star forming regions have
  been performed in recent years. They reached opposite conclusions
  concerning possible binary excesses in some of these associations.
  I develop a consistent method to reanalyze all these studies, so
  that I can compare all data consistently, and understand the
  previous findings. I also report the detection of five new
  companions to Taurus members.
  
  It appears that binary fraction in Taurus exceeds the main sequence
  value by a factor of 1.7 in the range 4--2000\,AU. The companion
  star fraction in this separation range is the same as the {\it
    overall\/} main sequence fraction. Ophiuchus, Chameleon, and
  possibly Lupus show similar excesses, although with lower confidence
  levels. Binaries in Ophiuchus seem to have larger flux ratios
  (towards faint companions) than in Taurus.
  
  It appears very unlikely that all very young star forming regions
  have binary excesses. The binary fraction seems to be established
  after $\sim1$\,Myr, but the precise nature of the difference between
  various regions is still unclear (overall binary fraction, orbital
  period distribution). It is not currently possible to put
  constraints on the binary formation models: higher angular
  resolution and larger sample sizes will be required.
  
\end{abstract}

\section{Introduction}

Surveys for low-mass main sequence (MS) binaries have poin\-ted out
that multiple system are numerous (53\% for G dwarfs,
Duquennoy~\&~Mayor \cite{dm}, hereafter DM). Only three years after
DM's paper, it was discovered that the Taurus star forming region
(SFR) had a very large number of binaries (Leinert {\it et al.\/}
\cite{leinert}, Ghez {\it et al.\/} \cite{ghez93}), many more than
expected, while the Orion Trapezium was ``normal'', as far as its
binary fraction is concerned (Prosser {\it et al.\/} \cite{prosser}).
Why is there such a difference between the two best studied SFR?
Taurus or the Trapezium could be exceptions, but it could also give
strong constraints on binary formation models.

What about the other well-known SFRs? Some studies have pointed out
binary excesses in Ophiuchus, Chameleon, Lupus and even in the
Trapezium cluster (Reipurth~\&~Zinnecker \cite{rz}, Ghez {\it et
  al.\/} \cite{ghez97}, Padgett {\it et al.\/} \cite{padgett}), but
some authors have also concluded that there was no excess in the same
regions (Simon {\it et al.\/} \cite{simon95}, Brandner {\it et al.\/}
\cite{b96}, Petr {\it et al.\/} \cite{petr}). Based on some of these
results, Ghez {\it et al.\/} (\cite{ghez93}) and Patience {\it et
  al.\/} (\cite{patience}) have proposed that the binary fraction was
decreasing with time, from high excess to MS values, while Bouvier
{\it et al.\/} (\cite{bouvier}) propose that the main factor driving
the binary's properties are the physical condition of the parent
cloud.
  
This paper intends to clarify previous results by analyzing all
available sets of data with a single consistent method allowing for
meaningful comparisons (Sect.\,\ref{method} and \ref{details}). I also
review the results on the older Hyades and Pleiades stars (Patience
{\it et al.\/} \cite{patience}, Bouvier {\it et al.\/}
\cite{bouvier}). Finally, an attempt to find a global trend will be
presented (Sect.\,\ref{results}).

\section{Method and hypotheses}
\label{method}

The main purpose of this work is to allow for direct comparisons
between all previous papers. Up to now, there are about as many
observational techniques as papers, and each has its own limitations.
Here, I will compare the results for each SFR to that of the MS, and
{\it then}, a comparison of the results with each other will be
meaningful; this method slightly increases the error bars by
accounting twice for the MS uncertainties, but it is usually a small
value.

As the largest low-mass MS survey is DM, only stars later than F7 on
the MS will be considered here. This corresponds to an early-G
spectral type for a 1 Myr star.  Thus, A and F stars will be excluded
from all SFR surveys. To take into account all companions, I calculate
the {\it companion star fraction}, which is the number of companions
per primary ($csf=\frac{B+2T+3Q}{S+B+T+Q}$, where $S$, $B$, $T$ and
$Q$ are the numbers of single, binary, triple and quadruple systems;
see Patience {\it et al.\/} \cite{patience}), instead of the {\it
  binary fraction\/} ($bf=\frac{B+T+Q}{S+B+T+Q}$); also, DM only
estimate $csf$. Finally, for each study, I tried to choose a
separation range over which the sensitivity is high enough so that all
companions can be found. These ranges are presented in Appendix~A.

The intuitive way of comparing all clusters is to select a wide
separation range and to count the number of binaries in this range for
each region. Due to the different distances involved, however, the
common separation range is narrow and leads to small numbers of
companions. The statistical significance of the results is thus quite
low.

The most powerful comparison of all clusters is obtained by comparing
each {\it csf\/} to that of the MS in the same separation range; each
SFR, however, has been surveyed in different ranges by different
studies. I calculate the total {\it csf\/} of the SFR from all surveys
and, concerning the MS value, I estimate it in the same separation
range for each survey (by integration of the analytic period
distribution given by DM) and I average these values using the number
of targets as a weight. In Table\,\ref{final}, $csf_{MS}$ is the
averaged MS value to be compared with the $csf$ in the 4th
column. $N_{\mathrm{obs}}$ is the total number of targets.

Two non critical assumptions are made: the total system mass is
1\,$M_\odot$ and the actual semi major-axis $a$ is linked to the
apparent separation $\rho$ via $\overline{\log a} = \overline{\log
  \rho} + 0.1$ (Reipurth~\&~Zinnecker \cite{rz}); reasonable changes
in these parameters does not change the {\it csf\/} by more than 1\%.

Also, some corrections have been applied in some cases (see
Sect.\,\ref{details}) to take into account poor and non homogeneous
dynamic range or selection biases. Concerning the dynamic range, it
has to be large enough to detect binaries with mass ratio
$q=0.1$\footnote{for systems with $M_1\sim 0.5M_\odot$, this limit
  would be well within the brown drawfs domain and this could modify
  the results.}, which is the lower limit of DM's survey (they cannot
observe binaries with $q<0.3$ for all targets, but they estimate a
correction down to this limit). Using Baraffe {\it et al.\/}
(\cite{isa})'s mass-luminosity relation at 2\,Myr, it appears that such
a mass ratio is equivalent to $\Delta K=2.9$, $\Delta I=3.6$ and
$\Delta V=4.3$\,mag at this age. These limits are reached in all
pre-main sequence (PMS) surveys except when a speckle technique is
used (these studies are limited to absolute magnitude and not flux
ratios, so that some stars were observed with worse sensitivities); in
this case, a correction has been applied to take into account the
strongly non uniform sensitivity of the survey.  All companions with
$q<0.1$ were excluded from the statistics to allow significant
comparisons with DM; this has not been done in the previous studies.
Determining mass ratios from flux ratios is somewhat hazardous for PMS
stars because of possible infrared excesses and different ages, but I
assume that this does not lead to any systematic bias. In older
clusters, as the mass-luminosity relation steepens with increasing
time, the uncompleteness correction becomes important, and cannot be
neglected for the Hyades and the Pleiades.

\section{Details of the calculation by clusters}
\label{details}

Table\,\ref{final} presents all the results developed in this section.
For each SFR, I explain what has been done (if anything) after simply
collecting the data from the literature.

\subsection{Taurus}
\label{taurus}

The speckle results of Ghez {\it et al.\/} (\cite{ghez93}) does not
need any correction, as all stars were observed with a large enough
dynamic range. The lunar occultation survey of Simon {\it et al.\/}
(\cite{simon95}), also reporting results from Richichi {\it et al.\/}
(\cite{richichi94}), however, suffers from a poor dynamic range, and I
applied a correction similar to Ghez {\it et al.\/} (\cite{ghez93}).
This method takes into account the fact that all stars were not
surveyed with the same dynamic range: the targets are binned by
relative brightness of approximately equal magnitude steps, and the
number of detected binaries in each bin is rescaled to the total
number of targets. Here, it adds $\sim4$ companions. The final
uncertainties are estimated from Poisson statistics on the {\it
  observed\/} number of companions and corrected {\it for each flux
  ratio bin} (this method gives a conservative estimation of the
error), and not on the final, corrected number of companions.

Recent $HST$ and adaptive optics images of the binary system HK Tau
have revealed a circumstellar disk around the secondary (Stapelfeldt
{\it et al.\/} \cite{hktau}). As it is seen edge-on, the star is
totally hidden, and we can only see scattered light. This explain why
$\Delta H=3.1$\,mag while the mass ratio is estimated to be about
$q\sim0.5$ from the spectral types of both components (Stapelfeldt
{\it et al.\/}). This system has not been excluded here. Otherwise,
three faint companions had to be excluded from the Leinert {\it et
  al.\/} (\cite{leinert}) survey. In some cases, Leinert {\it et
  al.\/} (\cite{leinert}) report the imaging results from
Reipurth~\&~Zinnecker (\cite{rz}) without further high angular
resolution observations. New images with adaptive optics have revealed
new subarcseconds companions to four of these systems (see
Appendix~B), which were added in this study.

\subsection{Ophiuchus}

Both Ghez {\it et al.\/} (\cite{ghez93}) and Simon {\it et al.\/}
(\cite{simon95}) results were corrected for incompleteness with the
same method as in Taurus, leading to an estimation of $\sim4$ missed
companions.

\begin{table*}[t]
\caption{\label{final}Comparison of star forming regions and MS
  samples regarding the companion star fraction. $^{{\mathrm
  \dagger}}$ the distance to the Orion complex is an average value.}
\begin{center}
\begin{tabular}{p{1.9truecm}p{0.5truecm}p{0.8truecm}ccccc|cl}
\hline
\vspace*{0.1truecm}

 & $N_{obs}$ & $N_{comp}$ & {\it csf} (1\,$\sigma$) & {\it csf}$_{\mathrm
 M\mathrm S}$ (1\,$\sigma$) & $\log\frac{csf}{csf_{\mathrm{MS}}}$ &
 statistical & ref. & distance & references \\
 & & & [\%] & [\%] & & significance & & [pc] & for distance \\
\hline
\multicolumn{10}{c}{PMS associations and clusters}\\
\hline
Tau-Aur & 117 & 67.1 & 57(8) & 34(4) & {\bf 0.22$\pm$0.08} & 2.8$\,\sigma$
 & 1,2,3,4 & 140 & Elias (1978) \\
Oph & 114 & 35.0 & 31(5) & 20(3) & {\bf 0.19$\pm$0.10} & 1.9$\,\sigma$ &
 2,3,4 & 160 & Chini (1981) \\
Oph & 95 & 24.6 & 26(5) & 13(2) & {\bf 0.30$\pm$0.11}
 & 2.7$\,\sigma$ & 2,4 & \hspace*{0.1 truecm}-- & \hspace*{0.7
 truecm}-- \\ 
Trapez. & 291 & 31 & 11(2) & 14(2) & {\bf -0.10$\pm$0.10} & 1.0$\sigma$ &
 5 & 450 & 5,6,7$^{{\mathrm \dagger}}$ \\
Trapez. & 34 & 2 & 6(4) & 8(1) & {\bf -0.12$\pm$0.29} & $<1\sigma$ & 6 &
 \hspace*{0.1 truecm}-- & \hspace*{0.2 truecm}-- \\
outer Trapez. & 50 & 6.6 & 13(5) & 10(1) & {\bf 0.11$\pm$0.17} &
 $<1\sigma$ & 7 & \hspace*{0.1 truecm}-- & \hspace*{0.2 truecm}-- \\
NGCs Ori & 99 & 12.4 & 13(4) & 10(1) & {\bf 0.10$\pm$0.15} & $<1\sigma$ &
 7 & \hspace*{0.1 truecm}-- & \hspace*{0.2 truecm}-- \\
Cha I & 85 & 19.2 & 23(6) & 16(2) & {\bf 0.16$\pm$0.13} & 1.2$\,\sigma$ &
 4,8,9 & 140 & Schwartz (1991) \\
Cha II & 23 & 5.0 & 22(10) & 13(2) & {\bf 0.23$\pm$0.21} & 1.1$\,\sigma$ &
 4,8,9 & 200 & Hughes \& Hartigan (1992) \\
Lup & 61 & 11.1 & 18(6) & 15(2) & {\bf 0.08$\pm$0.16} & $<1\,\sigma$
 & 4,8 & 150 & Krautter (1991) \\
CrA & 10 & 3.0 & 30(17) & 11(2) & {\bf 0.44$\pm$0.26} & 1.6$\,\sigma$
 & 4,8 & 130 & Marraco \& Rydgren (1981) \\
\hline
\multicolumn{10}{c}{{\it ROSAT\/} sources}\\
\hline
Tau-Aur & 68.6 & 25.1 & 37(12) & 26(4) & {\bf 0.15$\pm$0.16} &
 $<1\sigma$ & 10 & 140 & {\it assumed\/} \\
Cha & 86.8 & 4.4 & 5(3) & 7.5(1) & {\bf -0.21$\pm$0.31} & $<1\sigma$ & 9
 & 140 & \hspace*{0.4 truecm}-- \\
Sco-Lup & 64.4 & 7.5 & 12(7) & 7.5(1) & {\bf 0.20$\pm$0.26} &
 $<1\sigma$ & 9 & 150 & \hspace*{0.4 truecm}-- \\
\hline
\multicolumn{10}{c}{Older open clusters}\\
\hline
Pleiades & 144 & 40.8 & 28(6) & 27(3) & {\bf 0.02$\pm$0.10} & $<1\,\sigma$
& 11 & 130 & 11 \\
Hyades & 97 & 17.8 & 18(5) & 16(2) & {\bf 0.05$\pm$0.13} & $<1\,\sigma$ &
12 & 46.3 & Perryman {\it et al.\/} (1997) \\
\hline
\end{tabular}
\end{center}
References: 1 -- Leinert {\it et al.\/} (\cite{leinert}), 2 -- Ghez
{\it et al.\/} (\cite{ghez93}), 3 -- Simon {\it et al.\/}
(\cite{simon95}), 4 -- Reipurth~\&~Zinnecker (\cite{rz}), 5 -- Prosser
{\it et al.\/} (\cite{prosser}), 6 -- Petr {\it et al.\/}
(\cite{petr}), 7 -- Padgett {\it et al.\/} (\cite{padgett}), 8 -- Ghez
{\it et al.\/} (\cite{ghez97}), 9 -- Brandner {\it et al.\/}
(\cite{b96}), 10 -- K\"ohler~\&~Leinert (\cite{kl98}), 11 -- Bouvier
{\it et al.\/} (\cite{bouvier}), 12 -- Patience {\it et al.\/}
(\cite{patience}).
\end{table*}

\subsection{Orion clusters}

In the Trapezium, the results from Petr {\it et al.\/} (\cite{petr})
are uncorrected while Prosser {\it et al.\/} (\cite{prosser})
evaluated the number of unbound pairs (chance projection by crowded
fields). Here, I use their final results, where these non physical
pairs have been excluded.

In their study of the outer parts of the Trapezium and of NGC\,2024,
2068 and 2071, Padgett {\it et al.\/} (\cite{padgett}) evaluated the
probability for each companion to be a real companion. As they find
high {\it individual\/} probabilities, no correction is applied in the
study. The probability for {\it all companions\/} to be bound,
however, is rather small (55 and 62\% in the NGC clusters and the
Trapezium respectively), indicating that a correction is actually
needed. The average probabilities for each companion to be unbound are
4.0 and 6.2\% in the two subsamples. This is in agreement with the
averaged background companions probabilities given in Padgett {\it et
  al.\/}'s Table\,3: the predicted numbers of false detections are 0.6
and 0.4 respectively. I substracted these numbers in
Table\,\ref{final}, as well as two companions below the $q=0.1$ limit,
with increased error bars (Poisson statistics were applied to the
unbound pairs, too).

\subsection{Chameleon, Lupus, Corona Australis}

I applied the correction from Ghez {\it et al.\/} (\cite{ghez93},
\cite{ghez97}) to the limited subsample of low-mass stars in Ghez {\it
  et al.\/} (\cite{ghez97}), again with increased uncertainties. The
addition of the $csf$ from two independent subsamples, proposed by Ghez
{\it et al.\/}, leads to the same excess ratio as the method used
here.

\subsection{{\it ROSAT\/} population}

In all three SFRs, corrections for too faint companions ($q<0.1$) and
background projections are performed with the values given in Brandner
{\it et al.\/} (\cite{b96}) and K\"ohler \& Leinert (\cite{kl98}). The
third correction to apply is to take into account the bias induced by
the X-ray selection of the targets: a binary has two sources and can
thus be detected more easily in the {\it ROSAT\/} survey. I used a
similar method to Brandner {\it et al.} (I assume that the X-ray flux
of the secondary is independent of the primary's), but I replaced
their formula for $\Delta N$ by:
\[  \Delta N = \int_{0.5 L_{\mathrm{lim}}}^{L_{\mathrm{lim}}}
\rho(L_{1x})\int_{L_{\mathrm{lim}}-L_{1x}}^{L_{1x}} \rho^o(L_{2x})
{\mathrm d}L_{1x} {\mathrm d}L_{2x} \times bf \] with $\rho^o =
\frac{\rho}{\int_{L_{\mathrm{min}}}^{L_{1x}} \rho(L) {\mathrm d}L}$,
the normalized distribution of X-ray fluxes, {\it i.e.\/} the density
of probability for the secondary's flux; $L_{\mathrm{min}}=10^{21.5}
{\mathrm W}$ and $L_{\mathrm{max}}=10^{23.5} {\mathrm W}$ are the 
limits of validity for the flux distribution, and $L_{\mathrm{lim}}$
is the sensitivity limit of the X-ray surveys in each SFR. I then
applied a method similar to Brandner {\it et al.}, which both modifies
the sample size and the number of companions; I choosed $bf = 53$\%,
{\it i.e.\/} the main sequence value, but a value of 90\% does not
change the results in Table\,\ref{final} by more than 2\%. I find 9
faint, 4.3 background and 5.5 X-bias companions in Taurus; equivalent
figures in Chameleon and Scorpius are respectively (0, 1.0, 1.0) and
(1, 1.5, 3.0). The final number of targets is also a fractional number
($N_{\mathrm{obs}}=N-\Delta N$). If $x$ is the fraction of biased
binaries actually detected in the separation range of a survey, then
$N_{comp}=N_{comp,obs}-x\times\Delta N$. The corrections I evaluate
for Chameleon and Scorpius are smaller than that of Brandner {\it et
  al.}, because these were overestimations; in Taurus, the correction
estimated in K\"ohler \& Leinert is similar to that quoted in
Tab\,\ref{final}. Poisson uncertainties are associated to each
correction.

\subsection{Pleiades}

As already mentionned, a completeness correction is needed for this
cluster: from Henry \& McCarthy (\cite{hm}), $q=0.1$ corresponds to
$\Delta K=6$\,mag, which is not reached for all separations. The
assumption made by Bouvier {\it et al.\/} (\cite{bouvier}) is that the
DM's mass-ratio distribution can be used in the Pleiades, which seems
compatible with their results. The uncertainties, however, must be
increased, as Poisson statistics again apply to the observed numbers
of companions ({\it e.g.}, in the first bin of their Table\,2, the
total number of companions is $15\pm7.5$ and not $15\pm3.9$, since
they detect 4 companions).

\subsection{Hyades}
\label{hyades}

As in the Pleiades, a correction is needed, but it must be evaluated
{\it and\/} applied only to the subsample of low-mass stars (Patience
{\it et al.\/} \cite{patience} evaluate the correction on the whole
sample, but apply it to the low-mass stars). After excluding all stars
with spectral type earlier than F7, $M>1.25\,M_\odot$ and evolved stars
(see Tables\,2, 3 and 4 in Patience {\it et al.\/}; all stars with no
spectral type in their Table\,2 are excluded here), I checked that the
average detectable mass-ratio is unchanged ($q_{\mathrm{min}}=0.23$).
With the same correction as in the Pleiades, I estimate that 79\% of
the companions were detected in this survey. With a correction similar
to that proposed by Patience {\it et al.}, this number becomes 70\%;
the slight difference is due to the fact that Patience {\it et al.\/}
do not dismiss binaries with $q<0.1$. The 54\% reported in the
original study is due to the fact that a lot of higher mass stars
($M\sim2M\odot$) could hide many low-mass companions.

\section{Results and implications}
\label{results}

\subsection{Binary excesses in star forming regions}
\label{excess}

As already mentionned, the binary excess is strongly significant in
Taurus (99.5\% confidence level), where the overall binary fraction is
$\sim90$\% provided that the period distribution shape is the same as
the MS. Actually, the binary fraction given in Table\,\ref{final} is
comparable to the {\it overall\/} estimated {\it csf\/} in the MS
stars ($\sim 61\pm7\%$, DM). On the other hand, all studies of the
Orion clusters (Trapezium, NGC\,2024, 2068 and 2071) converge to a
``normal'' binary fraction. In the other SFRs, no obvious excess can
be detected, with the exception of Ophiuchus which is discussed below.
Although all data have been carefully analyzed, no definitive
conclusions can be drawn. Angular resolution still has to be improved
to increase the number of companions. The use of larger telecopes
equipped with adaptive optics, however, will not solve the main
problem. The low significance of the results is tightly linked to the
small sample sizes: except for the Trapezium cluster, there are always
less than 150 targets. Until an important embedded population is found
and surveyed, it will be very difficult to increase our confidence in
these results.

Reipurth \& Zinnecker (\cite{rz}) first proposed that there are more
binaries in PMS stars than in the MS. The excess they find is not
highly significant (about 1.5$\,\sigma$), but they use direct imaging,
without high angular resolution. Their sample consists mainly in
Ophiuchus, Chameleon and Lupus (213 out of 238 targets); here,
combining these three regions, the excess represents a factor of 1.5,
significant at the 2.6$\,\sigma$ level ({\it i.e.}, a probability of
99\% for these clusters to have a binary fraction different from the
MS).  Although each individual cluster does not contain enough stars,
this is an evidence that other SFRs than Taurus have binary excesses.
Actually, Ophiuchus and Chameleon both seem to have excesses
comparable to Taurus. If one excludes Simon {\it et al.\/}
(\cite{simon95}) data concerning Ophiuchus (see below), the excess
becomes a factor of 1.6, and the significance is increased to the
2.9$\,\sigma$ level.

Simon {\it et al.\/} (\cite{simon95}) do not find a binary excess in
Ophiuchus, while in Taurus, they end with a result similar to
Table\,\ref{final}. They point out the fact that they give only a
lower limit to the actual binary fraction, and the difference they
find seems to vanish in Table\,\ref{final} after averaging with the
results of Ghez {\it et al.\/} (\cite{ghez93}). The main difference
between Simon {\it et al.\/} survey and all other study of Ophiuchus
is that the former is sensitive to closer separations, thanks to a
lunar occulation technique. This could reveal a trend for Ophiuchus to
lack very close binaries ($\rho<0.1''$, below the limit of Ghez {\it
  et al.\/} \cite{ghez93} survey). However, Simon {\it et al.\/} find
31\% of their companions below 0.1$''$ in Ophiuchus and 27\% in
Taurus. Of course, these numbers suffer from poor statistics, but
there is no evidence for a difference in the period distributions in
their study.  Another possibility to explain the results of Simon {\it
  et al.\/} is the difference in flux ratios in Ophiuchus and Taurus.
From Ghez {\it et al.\/} (\cite{ghez93}), it appears that 73\% of the
binaries in Taurus have $\Delta K<1.5$\,mag, while only 23\% in
Ophiuchus have such flux ratios (the median flux ratios in both
samples are respectively $\Delta K\sim0.8$ and $\sim1.6$\,mag). From a
$\chi^2$ test, the probability that the two samples are drawn randomly
from the same distribution is smaller than 0.5\%. The median dynamic
range in Simon {\it et al.\/} survey is $\Delta K\sim1.7$\,mag, and it
is plausible that they do not find a binary excess in Ophiuchus
because they miss faint companions. The problem is then to understand
why the flux ratios are different in Ophiuchus and Taurus SFRs.

The {\it ROSAT\/} population is not easy to handle: as already
proposed by K\"ohler~\&~Leinert (\cite{kl98}), the ``X-sources'' in
Taurus are probably related to the molecular cloud since the binary
excess, although not statistically significant, is rather similar to
the other sources; this argument, however, is given {\it a fortiori},
and thus it is not very compelling. The resulting excess (a factor of
1.6) is significant at the 2.9$\,\sigma$ confidence level.  On the
other hand, in Chameleon, the X-ray selected population and the rest
of the cloud are different at the 1.2$\,\sigma$ level.  This is in
agreement with Brandner {\it et al.\/} (\cite{b96}), who find a
2$\,\sigma$ difference. However, they consider it as similar and
average the two results. It seems more likely that the {\it ROSAT\/}
population is not entirely related to the SFR, as proposed by
Neuh\"auser \& Brandner (\cite{nb98}), who find that 6 out of 7 stars
observed with {\it HIPPARCOS\/} are foreground stars. In
Scorpius-Lupus, finaly, the excess is large, although not very
significant. It is interesting to notice that, in the separation range
0.8--3$''$, Brandner {\it et al.\/} (\cite{b96}) find a very large
excess in Upper-Scorpius B and a MS result in Upper-Scorpius A (which
are two parts of the same SFR with different ages); the average of
these two values leads to the observed excess.  With high angular
resolution, Brandner \& K\"ohler (\cite{bk98}) proposed that the
period distribution are different in the two subclusters.

In older clusters (120 and 600\,Myr respectively for the Ple\-iades and
the Hyades), no strong binary excess is detected; the excess reported
by Patience {\it et al.\/} (\cite{patience}) is due to the large
correction they evaluated (see Sect.\,\ref{hyades}).

\subsection{Implications for binary formation}

As explained in Sect.\,\ref{excess}, Taurus is probably not the only
SFR with a large binary excess in the separation range
$\approx$10--2000 AU. Ophiuchus and Chameleon have probably similar
excesses, although the results are less significant than in Taurus. On
the other hand, several very young clusters (Trapezium, NGC clusters
in Orion) do not have such features. It can thus be excluded that {\it
  all\/} SFR have very high binary fractions at the begining of the
star formation stage: the possibility of a time evolution for all PMS
associations is very unlikely, at least after $\sim1$\,Myr, the typical
age of these SFR.

All the regions with a binary excess (Taurus, Chameleon, Ophiuchus)
are loose associations: all dense clusters have binary fractions
compatible with the MS. Unless the majority of the solar neighbourhood
stars were formed in Taurus-like associations, it appears that the
binary fraction does not evolve from the T Tauri phase to the MS. This
is an evidence for the low impact of gravitational encounters in {\it
  dense\/} clusters after 1\,Myr or so, and it is very unlikely that
such interactions can affect the binary fraction in loose SFR.
Furthermore, as the ratio of binaries to triple systems in PMS and MS
are similar (DM, Leinert {\it et al.\/} \cite{leinert}), it seems that
disruption of high order multiples due to unstable orbits are quite
rare too.

It is still unclear whether the {\it total binary fraction\/} is
higher in dense clusters or if the {\it period distribution\/} is
different, with more visual binaries and an overall fraction similar
to the MS.  In Taurus, however, the latter seems unlikely, as the
number of spectroscopic binaries is not extremely low (Mathieu
\cite{mathieu}).  In any case, it seems that the binary fraction in
the range $\approx10-2000$\,AU is established very soon in the history
of star formation, probably before $\sim1$\,Myr. Kroupa
(\cite{kroupa}) has shown, with $N$-body simulations, that wide
binaries may be massively disrupted in a time shorter than a few Myr,
provided it is extremely dense in its very first stages (it would
require $n\sim10^5$\,stars/pc$^3$).  Bate {\it et al.\/} (\cite{bate})
find a weak evidence that wide binaries ($a\gtrsim400$ AU) may have
actually been disrupted in the Trapezium cluster. At smaller
separations, however, it seems that no disruption has occured. It is
possible that all SFR start their evolution with a high binary
fraction and that the dense clusters disrupt most of the widest pairs.
However, it appears unlikely because no excess is found in Trapezium
down to $\sim50$\,AU, while it should still exist at such small
separations. Furthermore, the proportion of circumstellar disks is
very high in the Trapezium (Hillenbrand {\it et al.\/} \cite{lynne}),
implying that the number of encounters is not very high.

It is also possible that the binary fraction is established during the
formation process, without any later disruption. Durisen \& Sterzik
(\cite{durisen}) have pointed out that a natural prediction of both
cloud and disk fragmentation models is that the binary fraction is
higher in colder SFR. If the Trapezium and similar clusters had a high
initial gas temperature, it could explain the large excess of loose
associations with regard to the MS. On the other hand, Durisen {\it
  et al.\/} (in prep.) show that the cloud temperature could also
influence the orbital period distribution. This could account for the
results of Brandner \& K\"ohler (\cite{bk98}) who present evidence
that the peak of the orbital period distribution changes with the
physical conditions of the parent cloud; their numbers, however, are
very small, and their results need to be ascertained.

\section{Conclusion}

I developped a method with consistent corrections for uncompleteness
and selection biases to clarify the issue of the possible binary
excess in SFRs. Reanalyzing all published PMS binary surveys, it
appears that some previous conclusions have to be revisited, although
the major ones hold (Taurus and Ophiuchus SFR show binary excess,
while Orion and zero-age MS clusters do not). The following
conclusions are reached here:

\begin{itemize}
\item the Taurus members have a significant binary excess
  (2.8$\,\sigma$), with $\sim95$\% of stars being multiple systems if
  the orbital period distribution has the same shape as the MS
  binaries. I also report astrometry and near-infrared photometry for
  four new subarseconds companions.
\item combining Ophiuchus, Chameleon and Lupus, the excess is
  significant at the 2.6$\,\sigma$ confidence level at least, which
  increases the confidence in the result initially pointed out by
  Reipurth \& Zinnecker (\cite{rz}). It also appears that the binary
  infrared flux ratios are larger (towards fainter companions) in
  Ophiuchus than in Taurus.
\item all the Orion clusters, as well as the Pleiades and the Hya\-des,
  have binary fractions similar to the MS.
\item unlike in Taurus, there is no evidence from binarity that the
  $ROSAT$ populations in Chameleon and Lupus are linked to the clouds.
\end{itemize}

It is not currently possible to discriminate between a difference in
the overall binary fraction or in the orbital period distribution,
though the former appears more likely. The use of larger telescopes
with adaptive optics will be needed to settle this issue. Also, to
confirm and increase the confidence level of the binary excesses in
Ophiuchus or Chameleon, it will be necessary to survey a sample at
least twice as large as the current known population.

\begin{acknowledgements}
  I wish to thank J. Bouvier, F. M\'enard, J.-L. Mo\-nin, M. Bate and M.
  Simon for fruitful discussions and comments on early versions of
  this work. I also thank Ch. Leinert for his prompt and helpful referee
  report.
\end{acknowledgements}

\appendix
\section{Cluster distances and separation range of the different
  surveys}
\label{dist}

The distances assumed to convert angular to linear separations are
presented in Table\,\ref{final}. These values are not very accurate
(the uncertainties are probably about 20\%), but they are not
extremely important values, since the period distribution is rather
flat and, thus, a simple shift in the integration limits does not
change dramatically the results. Concerning the {\it ROSAT\/}
populations, I first assumed that they are linked to the SFRs, as
proposed by Guillout {\it et al.\/} \cite{guillout}, although this is
still unclear.

\begin{table}[bt]
\caption{\label{limits} Separation range for each study. ``diff'':
  diffraction limit; ``sens'': limit for homogeneous sensitivity;
  ``seeing'': worse seeing value; ``fov'': instrument field-of-view;
  ``back'': limit for small background stars contamination. References
  are the same as in Table\,\ref{final}.}
\begin{center}
\begin{tabular}{lccc}
\hline
Ref. & sep. range ($''$) & lower limit & upper limit \\
\hline
1 & 0.13--13.0 & diff & back \\
2 & 0.1--1.8 & sens & fov \\
3 & 0.02--10.0 & sens & back \\
4 & 1.0--12.0 & seeing & back \\
5 & 0.1--1.0 & sens & back \\
6 & 0.14--0.5 & diff & back \\
7 & 0.3--2.3 & sens & back \\
8 & 0.1--1.2 & sens & fov \\
-- & 1.2--12.0 & seeing & back \\
9 & 0.8--3.0 & seeing & back \\
10 & 0.13--13.0 & diff & back \\
11 & 0.08--7.0 & diff & fov \\
12 & 0.1--1.07 & sens & fov \\
\hline
\end{tabular}
\end{center}
\end{table}

More critical are the separation ranges in which each study was
performed (Table\,\ref{limits}). When the technique is simple imaging
(without high angular resolution), I choose the worst value of the
seeing as the lower limit, so that all stars were observed under
better or equivalent conditions. For speckle surveys, it is possible
to choose a lower limit such that the sensitivity is almost constant
for larger separations without loosing too much companions;
furthermore, the scatter in the individual sensitivities is so large
that it is impossible to apply a uniform correction to the whole
sample. Adaptive optics studies, on the other hand, are
diffraction-limited and very uniform in sensitivity, but the
sensitivity increases gradually up to roughly 1$''$, and it is not
interesting to use this wide lower limit; I just kept the diffraction
limit, although I acknowledge that a few companions may have been
missed. The dynamic range, however, is usually about 2\,mag around the
first Airy ring; the majority of the close companions are thus
detected at near-infrared wavelength. Binary separations can be
limited either by the instrument field-of-view or by the need to avoid
background contamination (this is discussed in each study), the latter
beeing a problematic issue.

\section{New companions in Taurus}
\label{new}

On December 11 and 18, 1997, near-infrared images of Taurus binaries
from Leinert {\it et al.\/} (1993) and Reipurth~\&~Zinnecker~(1993)
samples were taken at CFHT, using the adaptive optics system and the
new near-infrared camera, KIR. While obtaining $JHK$ photometry of
already known multiple systems, five new companions were discovered in
binary and triple systems. I report in Table \ref{newdata} the
astrometry and relative photometry; all details will be published in a
forthcoming paper (Monin {\it et al.}, in prep.). The uncertainties
for the astrometry are smaller than 2\% for the separation and 0.5$^o$
for the position angle (measured eastwards from the North); the
relative photometry is accurate to about 0.05 mag in each band.

The status of the faint star close to HBC\,358 is not addressed here,
although at this separation, the probability that this is a background
projected companion is low (the fact that the flux ratio decreases
with increasing wavelength, however, could be an evidence for its
background location). Given the flux ratio, anyway, it has not been
included in this paper.

\begin{table}[t]
\caption{\label{newdata} Astrometry and relative photometry for the 4
  new subarcsecond companions in Taurus. Astrometry and photometry for
  HBC 358 C are with regard to HBC 358 Aa (brightest component in $JHK$).}
\begin{center}
\begin{tabular}{lccccc}
\hline
primary & sep. ($''$) & PA ($^o$) & $\Delta J$ &  $\Delta H$ &
$\Delta K$ \\
\hline
UX Tau B & 0.138 & 303.9 & 0.29 & 0.28 & 0.27 \\
J 4872 A & 0.175 & 76.4 & 0.26 & 0.16 & 0.14 \\
Haro 6-37 A & 0.333 & 181.2 & 1.78 & 1.67 & 1.57 \\
HBC 358 A & 0.150 & 334.0 & 0.06 & 0.07 & 0.09 \\
\hline
HBC 358 C & 3.15 & 331.2 & 4.35 & 4.33 & 4.49 \\
\hline
\end{tabular}
\end{center}
\end{table}

\end{document}